\documentclass[aps,11pt,eqsecnum, preprint,nofootinbib,superscriptaddress]{revtex4}
\usepackage{amssymb,amsmath,amsthm,graphicx,amscd}
\usepackage{enumerate,color,changepage,ulem}
\usepackage[hidelinks]{hyperref}

\begin{document} 
\title{\Large Nonequilibrium Quantum Free Energy and Effective Temperature, Generating Functional and Influence Action} 

\author{Jen-Tsung Hsiang}
\email{cosmology@gmail.com}
\affiliation{Center for High Energy and High Field Physics, National Central University, Chungli 32001, Taiwan, ROC}
\author{Bei-Lok Hu}
\email{blhu@umd.edu}
\affiliation{Maryland Center for Fundamental Physics and Joint Quantum Institute, University of Maryland, College Park, Maryland 20742-4111, USA}

\begin{abstract}
A definition of nonequilibrium free energy $\mathcal{F}_{\textsc{s}}$ is proposed for dynamical Gaussian quantum open systems strongly coupled to a heat bath and a formal derivation is provided by way of the generating functional in terms of the coarse-grained effective action and the influence action. For Gaussian open quantum  systems exemplified by the quantum Brownian motion model studied here, a time-varying effective temperature can be introduced in a natural way, and with it, the nonequilibrium free energy $\mathcal{F}_{\textsc{s}}$, von Neumann entropy $\mathcal{S}_{vN}$ and internal energy $\mathcal{U}_{\textsc{s}}$ of the reduced system ($S$) can be defined accordingly. In contrast to the nonequilibrium free energy found in the literature which references the bath temperature, the nonequilibrium thermodynamic functions we find here obey the familiar  relation $\mathcal{F}_{\textsc{s}}(t)=\mathcal{U}_{\textsc{s}}(t)- T_{\textsc{eff}} (t)\,\mathcal{S}_{vN}(t)$ {\it at any and all moments of time} in the system's fully nonequilibrium evolution history. After the system equilibrates they coincide, in the weak coupling limit, with their counterparts in conventional equilibrium thermodynamics. Since the effective temperature captures both the state of the system and its interaction with the bath, upon the system's equilibration, it approaches a value slightly higher than the initial bath temperature. Notably, it remains nonzero for a zero-temperature bath, signaling the existence of system-bath entanglement. Reasonably, at high bath temperatures and under ultra-weak couplings, it becomes indistinguishable from the bath temperature. The nonequilibrium thermodynamic functions and relations discovered here for dynamical Gaussian quantum systems should open up useful  pathways toward  establishing meaningful theories of nonequilibrium quantum thermodynamics.   
\end{abstract}
\maketitle

\hypersetup{linktoc=all}

\newtheorem{theorem}{Theorem}
\baselineskip=18pt
\allowdisplaybreaks

\section{Introduction}

Using the techniques and language of quantum field theory (QFT) for the description of the statistical mechanics  of quantum many-body systems \cite{ZJ02} has a successful history of well over half a century. The first stage was dominated by imaginary-time finite temperature QFT~\cite{FetWal,KG06,MB96} which has proven its high utility for the study of equilibrium quantum systems in a vast range of fields.  Real-time formulation~\cite{LW87} and nonequilibrium Green functions~\cite{KadBay} are necessary for treating time-dependent quantum systems \cite{HK04}. The most powerful toolbox for the study of nonequilibrium (NEq) quantum systems in our opinion is the Schwinger-Keldysh close-time-path (CTP) formalism~\cite{CTP} which easily enables the use of diagrammatic techniques and the Feynman-Vernon influence functional (IF)~\cite{IF} formalism which is naturally adept to treating open quantum systems~\cite{UW12,BrePet,RivHue}. They contributed to the synergy between QFT and nonequilibrium statistical mechanics as witnessed by the fast advancement of nonequilibrium quantum field theory~\cite{CalHu88,CalHu08,JR09,AK11,Berges15} in the last three decades.

\subsection{Nonequilibrium quantum thermodynamic functions}

While it is reasonable to assume some theoretical confidence that equilibrium quantum thermodynamics can be obtained from nonequilibrium QFT as a limiting case after the system equilibrates -- for those systems which indeed can do so -- it remains a nontrivial challenge to come up with well-defined  thermodynamic functions under fully dynamical nonequilibrium conditions, and even more of a challenge, to show that the equilibrium thermodynamic relations also hold for these nonequilibrium thermodynamical potentials. On the practical side, to meet the needs of quantum sciences and technologies, establishing viable theories of quantum thermodynamics for the parameter regimes where the underlying assumptions of classical thermodynamics breaks down -- at low temperatures, for small systems, with sizable fluctuations and persistent memories --  becomes a pressing necessity.

In this work we shall focus on  nonequilibrium free energy, arguably the most important thermodynamic function, and also the most studied. Most important because many equilibrium thermodynamic quantities in canonical distribution can be obtained from the free energy.  In the statistical mechanics literature, for classical systems, Esposito and Van den Broeck~\cite{EspVdB} define (in their Eq.~(6)) the nonequilibrium system free energy as  $\mathcal{F}_{\textsc{s}}(t) =\mathcal{U}_{\textsc{s}}(t)-T_{\textsc{b}}\mathcal{S}_{vN}(t)$, (we have adapted it with notations used for easy comparison) where $T_{\textsc{b}}$ is the  temperature  of an ideal heat bath held \textit{constant} with which the system is in contact.  Deffner and Lutz~\cite{DefLut11} (in their Eq.~(7)) defines the NEq free energy by relating it to the relative (Kullback-Leibler) entropy between the nonequilibrium state $\hat{\rho}(t)$ of the system driven by an external agent $\alpha$ and the instantaneous equilibrium (canonical) distribution $\hat{\rho}_{eq}(t)$ corresponding to the system Hamiltonian that takes the same value of $\alpha$ at that moment. For a lucid description of nonequilibrium free energy for isothermal processes in quantum systems and related quantum information issues, we refer to the review of Parrondo, Horowitz, and Sagawa~\cite{PHS15} and references therein.

The so-called `nonequilibrium free energy' is most studied because of its linkage to the famous fluctuation theorems of Jarzynski~\cite{Jarzyn} and  Crooks~\cite{Crooks} (J-C). Many authors call the free energy in this context as `nonequilibrium free energy' referring to the difference between the free energies of the initial and final \textit{equilibrium} states of the system (on the RHS of the J-C relation) which is equal to and obtainable from the \textit{nonequilibrium work function} (on the LHS of the J-C relation). E.g., in Evans~\cite{Evans}, an elegant derivation of the Jarzynski-Crooks relation  valid for time reversible \textit{deterministic} systems is given, highlighting the close relationship between the non-equilibrium free energy theorems and the fluctuation theorem. Or, e.g., in Sivak and Crooks \cite{SivCro},   a simple near-equilibrium approximation for the free energy is found in terms of an excess mean time-reversed work, which can be experimentally measured on real systems. This is an important subject which we shall closely examine in a later work. Another noted difference between our setup of \textit{free} open systems (systems coupled to a heat bath but otherwise free of applied forces) and those in the fluctuation theorems context is the presence of an external agent doing work on the systems. E.g, Parrondo et. al.,~\cite{EspVdB} (Box 1, Eq.~(2)), define 
\begin{adjustwidth}{20pt}{20pt}
	``\textit{a non-equilibrium free energy for a generic statistical state $\hat{\rho}$ of a system, with a Hamiltonian $\hat{H}_0$, in contact with a thermal bath by
\begin{equation}
	\mathcal{F} (\hat{\rho}, \hat{H}_0) = \langle\hat{H}_0\rangle_{\hat{\rho}}- T_{\textsc{b}}\,\mathcal{S}_{vN}(\hat{\rho})
\end{equation}
where $\langle\hat{H}_0\rangle_{\hat{\rho}}$ is understood as $\langle\hat{H}_0\rangle_{\hat{\rho}}=\operatorname{Tr}_{\textsc{s}}\{\hat{\rho}\,\hat{H}_0\}$. The von Neumann entropy and the associated non-equilibrium free energy are analogous to their equilibrium counterparts in non-equilibrium isothermal processes. Here, isothermal implies that the system is in contact with a thermal reservoir at \textit{constant} temperature $T_{\textsc{b}}$, although the system itself may not have a well-defined temperature. The minimal work, on average, necessary to isothermally drive the system from one arbitrary state to another is simply the difference, $\Delta\mathcal{F}$ between the non-equilibrium free energy in each state. The excess work with respect to this minimum is the dissipated or irreversible work}" contributing to entropy increase.
\end{adjustwidth}

\noindent Note the presence of an external agent doing work on their system, driving it through a nonequilibrium evolution, while the free energy difference there refers to the free energy of their system in two different states, both are under thermal \textit{equilibrium} conditions. Whereas in our setup the free system is coupled with arbitrary strength with the same heat bath and we seek to define a mathematically well-defined and physically meaningful NEq free energy and  NEq temperature for the system throughout its evolution in time.

\subsection{Quantum field theoretical approaches}

In terms of QFT approach to this issue  we mention early work of Berges and Wetterich~\cite{BerWet} for equilibrium conditions and \'Eboli, Jackiw and Pi~\cite{EJPi}, who use a variational principle to derive a Liouville-von Neumann equation for quantum fields out of equilibrium. In a functional Schr\"odinger picture they produce the isentropic, but energy-nonconserving, time evolution of mixed quantum states. Two more recent  papers by van Zon et al. \cite{vanZon} use nonequilibrium path integrals to calculate the free energy difference for quantum systems. However, since they use the imaginary-time path-integral representation of the canonical partition function and  introduce fictitious Hamiltonian dynamics it is not a \textit{bona fide} treatment of nonequilibrium physical processes. As the authors stated, there is significant difference between the fictitious dynamics and real quantum dynamics, even though the free energy found from their nonequilibrium method with fictitious dynamics is the exact quantum free-energy difference in the Jarzynski relation. The work computed in their scheme bears no relation to the work performed in any real quantum process except in the classical limit.   

Turning now to our present work,  as mentioned earlier we aim at arriving at a free energy well defined at all times throughout the reduced system's evolution in the open-system framework. To make transition from imaginary-time finite temperature QFT, we begin with the partition function under equilibrium conditions. Formulation of thermal field theory in terms of effective potentials has a long history. We then highlight its correspondence with a real-time formulation. Both formulations can be fused into a unified language in an open quantum system framework. When the influence of the thermal bath field need be accounted for, albeit in an averaged or coarse-grained manner, one can use the \textit{coarse-grained effective action}~\cite{JH1,CalHu08}, which contains the Feynman-Vernon influence action~\cite{IF,HPZ}, itself being the CTP effective action~\cite{CTP} of the bath field. The reduced density matrix elements then can be expressed as the forward- and backward-time path integrals weighed by coarse-grained effective action. The trace of this reduced density matrix with a proper normalization then gives us the generalized generating function or \textit{nonequilibrium partition function}. One key issue in this formulation is to identify the suitable normalization factor~\cite{TK16,TH20}.

\subsection{Key points and major findings}


The model we use for this investigation is the vintage quantum Brownian motion of a harmonic oscillator coupled with arbitrary strength bilinearly to a scalar field thermal bath~\cite{HPZ,GWT84,HWL08,UW12,BrePet}. We begin in Sec.~\ref{S:ohtbfgd} with a short summary of the imaginary and real time description of thermal field theory, in so doing transcribing the major statistical thermodynamic  quantities and relations to a quantum field theoretic formulation, highlighting their correspondences such as that of the partition function to the generating functional, the free energy to the effective action. Our system is initially placed not in a thermal state, but in an arbitrary Gaussian state, thus out of equilibrium from the bath. After the system begins to interact with the bath the full effect of the environment on the system is registered in the covariance matrix elements. In Sec.~\ref{E:gbejtejsd}, we seek the generic operator form of the density matrix of this open quantum system. For this we use the displacement, squeeze and rotation parametrization which can be expressed in terms of the time-dependent covariance matrix elements. 

Invoking the McCoy theorem in this construction we can uniquely relate the nonequilibrium parameter $\vartheta$ and the squeeze parameter $\eta$ with the covariance matrix elements associated with the density operator of the reduced quantum system and identify the \textit{nonequilibrium partition function}, Eq.~\eqref{E:gjhdfsjwe} for the reduced system. This is the first important result of this paper. For the Gaussian system under study, squeezing is a consequence of the system-bath interaction. It remains in nonequilibrium until equilibration at late times, after which the nonequilibrium partition function, in the weak coupling limit, becomes the familiar equilibrium partition function. With this nonequilibrium partition function we can identify in Sec.~\ref{S:gbseruhds} a time-dependent parameter $\beta_{\textsc{eff}}$ in \eqref{E:fgrhjss} which serves as a \textit{nonequilibrium effective temperature}. With this, it is natural to identify the nonequilibrium free energy, Eq.~\eqref{E:rbghdsfg}. These are the second and third important results in our findings.

Having fleshed out the essential ingredients sought after, we then proceed in Sec.~\ref{S:gbseruhds} to understand the derivations of these results at the formal level in quantum field theory, knowing that the statistical thermodynamics under the equilibrium conditions are well anchored in the path integral formulation. Indeed, by way of the closed-time-path formalism we can describe the nonequilibrium quantum thermodynamics of this system. Using the time-varying effective temperature $T_{\textsc{eff}}(t)$ the nonequilibrium free energy is shown to be related to  the coarse-grained effective action which contains the influence action. It obeys the familiar thermodynamic relation, now extended to nonequilibrium conditions: $\mathcal{F}_{\textsc{s}}(t)=\mathcal{U}_{\textsc{s}}(t)- T_{\textsc{eff}} (t)\,\mathcal{S}_{vN}(t)$  where $\mathcal{U}_{\textsc{s}}$ is the system's internal energy and $\mathcal{S}_{vN}$ is the von Neumann entropy of the reduced system. This is the fourth important result in our findings and arguably a distinctly attractive feature, namely, that the nonequilibrium partition function, free energy, internal energy, and entropy have the same functional dependence on the \textit{effective} temperature as their counterparts in the weak-coupling thermodynamics on the \textit{system} temperature. In this identification, the Hamiltonian of mean force arises~\cite{HTL11} and enters in strong coupling thermodynamics~\cite{GT09,Se16,JA17,SE17,QTDpot,Rivas}. Recently, alternative constructions for the Hamiltonian of mean force have been proposed in~\cite{Rivas,SE20,TH20a}.  In Sec.~\ref{S:rbtkfhskrt}, more discussions on the nonequilibrium effective temperature will be found. Further developments based on the results here are contained in two companion papers, one on the comparison of the system's internal energy and the Hamiltonian of mean force~\cite{HMF}, the other on entropy, entanglement and the second law~\cite{QTD2}. 


\section{partition function in equilibrium thermodynamics}\label{S:ohtbfgd}

We first consider a quantum harmonic oscillator, with mass $m$ and oscillating frequency $\omega$, in thermal equilibrium at temperature $\beta^{-1}$ in the framework of conventional weak-coupling thermodynamics. The partition function $\mathcal{Z}_\beta$ can be readily found by~\cite{FR65}
\begin{equation}\label{E:fkgbrtss}
	\mathcal{Z}_\beta=\operatorname{Tr}e^{-\beta\hat{H}_\beta}=\frac{1}{2}\,\operatorname{csch}\frac{\beta\omega}{2}\,,
\end{equation}
with the Hamiltonian taking the form
\begin{equation}
	\hat{H}_\beta=\frac{\hat{p}^{2}}{2m}+\frac{m\omega^{2}}{2}\,\hat{\chi}^{2}\,.
\end{equation}
in which $\hat{\chi}$ is the displacement operator of the oscillator and $\hat{p}$ is the conjugated momentum operator, satisfying $[\hat{\chi},\hat{p}]=i$. The equilibrium free energy $\mathcal{F}_\beta$, which is equivalent to the generating functional of the connected thermal correlation functions in the context of the functional method, is then given by
\begin{equation}\label{E:bgruwia}
	\mathcal{F}_\beta=-\frac{1}{\beta}\,\ln\mathcal{Z}_\beta=\frac{\omega}{2}+\frac{1}{\beta}\,\ln\bigl(1-e^{-\beta\omega}\bigr)\,.
\end{equation}
The position uncertainty $\langle\hat{\chi}^{2}\rangle$ and the momentum uncertainty $\langle\hat{p}^{2}\rangle$ are respectively given by
\begin{align}
	\langle\hat{\chi}^{2}\rangle&=\frac{1}{2m\omega}\,\coth\frac{\beta\omega}{2}\,,&\langle\hat{p}^{2}\rangle&=\frac{m\omega}{2}\,\coth\frac{\beta\omega}{2}\,,
\end{align}
since in the thermal state of the oscillator we have $\langle\hat{\chi}\rangle=0=\langle\hat{p}\rangle$. We immediately find the mean values of the energy is
\begin{equation}
	\langle\hat{H}_\beta\rangle=\frac{\omega}{2}\,\coth\frac{\beta\omega}{2}=-\frac{\partial}{\partial\beta}\ln\mathcal{Z}_\beta\,.
\end{equation}
Here in taking the averages we trace over the complete set of the Fock number states. In the context of the open systems, it is more convenient to use the eigenstates of the displacement operator.

Thus, to prepare for a comparison with the ensuing treatment of open Gaussian systems, we will re-formulate the previous results. In analogy to the unitary evolution operator, we find
\begin{equation}\label{E:roytoyer}
	\langle\chi\vert e^{-\beta\hat{H}_\beta}\,\vert\chi'\rangle=\biggl(\frac{m\omega}{2\pi\sinh\beta\omega}\biggr)^{\frac{1}{2}}\exp\biggl\{-\frac{m\omega}{2\sinh\beta\omega}\biggl[\bigl(\chi^{2}+\chi'^{2}\bigr)\cosh\beta\omega-2\chi\chi'\biggr]\biggr\}\,.
\end{equation}
It explicitly shows that the matrix elements of an exponentiated quadratic operator yields not only a Gaussian function of $\chi$ and $\chi'$, but contains an additional factor. When we attempt to generalize the current result to a general Gaussian state, it can be nontrivial to identify the corresponding factor from the matrix elements of the reduced density matrix.

Eq.~\eqref{E:roytoyer} implies
\begin{equation}\label{E:fkhbkfgsf}
	\mathcal{Z}_\beta=\int_{-\infty}^{\infty}\!d\chi\int_{-\infty}^{\infty}\!d\chi'\;\delta(\chi-\chi')\,\langle\chi\vert e^{-\beta\hat{H}_\beta}\,\vert\chi'\rangle=\frac{1}{2}\,\operatorname{csch}\frac{\beta\omega}{2}\,,
\end{equation}
such that the corresponding density matrix element of the oscillator in its thermal state is given by
\begin{align}\label{E:gnrtsgls}
	\langle\chi\vert\,\hat{\rho}_{\beta}\,\vert\chi'\rangle&=\frac{1}{\mathcal{Z}_\beta}\,\langle\chi\vert e^{-\beta\hat{H}_\beta}\,\vert\chi'\rangle\notag\\
	&=\biggl(\frac{m\omega}{\pi}\,\tanh\frac{\beta\omega}{2}\biggr)^{\frac{1}{2}}\exp\biggl\{-\frac{m\omega}{2\sinh\beta\omega}\biggl[\bigl(\chi^{2}+\chi'^{2}\bigr)\cosh\beta\omega-2\chi\chi'\biggr]\biggr\}\,.
\end{align}
We then readily recover the earlier results
\begin{align}
	\langle\hat{\chi}^{2}\rangle&=\int_{-\infty}^{\infty}\!d\chi\int_{-\infty}^{\infty}\!d\chi'\;\delta(\chi-\chi')\,\chi^{2}\langle\chi\vert\,\hat{\rho}_{\beta}\,\vert\chi'\rangle=\frac{1}{2m\omega}\,\coth\frac{\beta\omega}{2}\,,\\
	\langle\hat{p}^{2}\rangle&=\int_{-\infty}^{\infty}\!d\chi\int_{-\infty}^{\infty}\!d\chi'\;\delta(\chi-\chi')\biggl[-\frac{\partial^{2}}{\partial\chi^{2}}\langle\chi\vert\,\hat{\rho}_{\beta}\,\vert\chi'\rangle\biggr]=\frac{m\omega}{2}\,\coth\frac{\beta\omega}{2}\,.
\end{align}
Relating the density matrix elements \eqref{E:gnrtsgls} with the general expressions of the Gaussian state, we will see an interesting connection between the partition $\mathcal{Z}_\beta$ and the the uncertainty principle.

The matrix elements $\rho(\chi,\chi')=\langle\chi\vert\,\hat{\rho}\,\vert\chi'\rangle$ of a general Gaussian state in the $\chi$-representation can be written as
\begin{align}\label{E:nneker}
	\rho(\chi,\chi')&=\biggl(\frac{a_{1}+a_{2}+a_{3}}{\pi}\biggr)^{\frac{1}{2}}\,\exp\biggl[-\frac{(a_{4}+a_{5})^{2}}{4(a_{1}+a_{2}+a_{3})}\biggr]\notag\\
	&\qquad\qquad\qquad\qquad\times\exp\Bigl[-\bigl(\alpha_{1}\,\Sigma^{2}+\alpha_{2}\,\Sigma\Delta+\alpha_{3}\,\Delta^{2}+\alpha_{4}\,\Sigma+\alpha_{5}\,\Delta\bigr)\Bigr]\,,
\end{align}
with
\begin{align*}
	\alpha_{1}&=+\frac{1}{2\langle\Delta\hat{\chi}^{2}\rangle}\,,&\alpha_{2}&=-i\,\frac{\langle\{\Delta\hat{\chi},\Delta\hat{p}\}\rangle}{2\langle\Delta\hat{\chi}^{2}\rangle}\,,\\
	\alpha_{3}&=+\frac{\langle\Delta\hat{p}^{2}\rangle}{2}-\frac{\langle\{\Delta\hat{\chi},\Delta\hat{p}\}\rangle^{2}}{8\langle\Delta\hat{\chi}^{2}\rangle}\,,&\alpha_{4}&=-\frac{\langle\hat{\chi}\rangle}{\langle\Delta\hat{\chi}^{2}\rangle}\,,\\
	\alpha_{5}&=-i\,\langle\hat{p}\rangle+i\,\frac{\langle\hat{\chi}\rangle}{2\langle\Delta\hat{\chi}^{2}\rangle}\,\langle\{\Delta\hat{\chi},\Delta\hat{p}\}\rangle\,,\\
	a_{1}+a_{2}+a_{3}&=\frac{1}{2\langle\Delta\hat{\chi}^{2}\rangle}\,,&\frac{(a_{4}+a_{5})^{2}}{4(a_{1}+a_{2}+a_{3})}&=\frac{\langle\hat{\chi}\rangle^{2}}{2\langle\Delta\hat{\chi}^{2}\rangle}\,,
\end{align*}
with $\Sigma=(\chi+\chi')/2$, $\Delta=\chi-\chi'$ and $\Delta\hat{\chi}=\hat{\chi}-\langle\hat{\chi}\rangle$. The thermal state is a special case of the Gaussian state, with $\alpha_{2}=\alpha_{4}=\alpha_{5}=0$, and 
\begin{align}\label{E:bgkdvgshsbs}
	\alpha_{1}&=+\frac{1}{2\langle\hat{\chi}^{2}\rangle}\,,&\alpha_{3}&=+\frac{\langle\hat{p}^{2}\rangle}{2}\,,&a_{1}+a_{2}+a_{3}&=\frac{1}{2\langle\hat{\chi}^{2}\rangle}\,.
\end{align}
Comparing with \eqref{E:nneker}, we find \eqref{E:roytoyer} can be expressed as
\begin{align}\label{E:fgkrugyvs}
	\langle\chi\vert e^{-\beta\hat{H}_\beta}\,\vert\chi'\rangle=\biggl[\frac{1}{\pi}\biggl(\alpha_{3}-\frac{\alpha_{1}}{4}\biggr)\biggr]^{\frac{1}{2}}\,\exp\biggl\{-\biggl(\alpha_{3}+\frac{\alpha_{1}}{4}\biggr)\bigl(\chi^{2}+\chi'^{2}\bigr)+2\biggl(\alpha_{3}-\frac{\alpha_{1}}{4}\biggr)\chi\chi'\biggr\}\,,
\end{align}
with
\begin{align}\label{E:zgxmeowe}
	\alpha_{1}&=m\omega\,\tanh\frac{\beta\omega}{2}\,,&\alpha_{3}&=\frac{m\omega}{4}\,\coth\frac{\beta\omega}{2}\,.
\end{align}
Using \eqref{E:fkhbkfgsf} and \eqref{E:fgkrugyvs}, we find that the partition function $\mathcal{Z}_\beta$ is given by
\begin{equation}\label{E:dfkhsbcx}
	\mathcal{Z}_\beta=\Bigl(\frac{\alpha_{3}}{\alpha_{1}}-\frac{1}{4}\Bigr)^{\frac{1}{2}}=\Bigl[\langle\hat{\chi}^{2}\rangle\langle\hat{p}^{2}\rangle-\frac{1}{4}\Bigr]^{\frac{1}{2}}
\end{equation}
after substituting \eqref{E:bgkdvgshsbs} in \eqref{E:dfkhsbcx}. Eq.~\eqref{E:fkgbrtss} is recovered with the use of \eqref{E:zgxmeowe}. However, interestingly, it  reveals a hidden structure, namely the partition function \eqref{E:dfkhsbcx} contains the uncertainty relation,
\begin{equation}\label{E:nfgjits}
	\langle\hat{\chi}^{2}\rangle\langle\hat{p}^{2}\rangle\geq\frac{1}{4}\,.
\end{equation}
We shall show that for more general cases the Schr\"odinger-Robertson uncertainty relation emerges when there are cross correlations between the canonical variables in \eqref{E:nfgjits}.

Systems in an equilibrium thermal state can alternatively be described in terms of the Feynman path integral formalism~\cite{HK04,MB96,ZJ02,KG06} in which the partition function $\mathcal{Z}_\beta$ is given by a path integral along the imaginary time $\tau$, from $\tau=0$ to $\tau=\beta$
\begin{equation}\label{E:dkbshfgbd}
	\mathcal{Z}_\beta=\operatorname{Tr}_\beta e^{-\beta\hat{H}_\beta}=\int\mathcal{D}\chi\;e^{-S_{E}[\chi,\dot{\chi}]}
\end{equation}
where $S_{E}$ is the Euclidean action of the system corresponding to the Hamiltonian $\hat{H}_\beta$ and $\dot{\chi}$ in this case denotes the derivative of the system's canonical variable $\chi$ with respect to the imaginary time $\tau$. It has a simple connection\footnote{Subtleties may arise when we perform analytic continuations between the real-time and the imaginary-time formalism, and when we identify the suitable Green's functions in the context. Detailed discussions can be found in~\cite{HK04,MB96,ZJ02,KG06}.} with the real-time path integral formalism for the transition amplitude of the system undergoing an unitary evolution, namely, if we perform a Wick rotation of time from its real axis to the imaginary axis, i.e., $t=-i\,\tau$ with $\tau\in\mathbb{R}$. Then the generating functional in the real-time formalism will be analogous to the partition function in statistic mechanics. This correspondence persists even when we include the external $c$-number sources in our system, thus facilitating the construction of Green's functions. An important thermodynamic quantity, the free energy, $\mathcal{F}_\beta=-\beta^{-1}\,\ln \mathcal{Z}_\beta$, is formally the counterpart of the effective action $W$, or the generating functional of the connected Green's functions, in the real- and imaginary-time path integral analogy~\cite{LW87}. Moreover, both formalisms can be nicely unified under the close-time path integral formalism~\cite{CTP,IF}, which, in the framework of quantum open systems, is particularly suitable for the description of the dynamical evolution of nonequilibrium interacting systems~\cite{CalHu08,JR09,AK11}. This is the pathway we shall take to generalize the concept of free energy from an equilibrium setting to a nonequilibrium scenario. In the next section, we will formulate the nonequilibrium partition function and free energy for Gaussian open systems, as a generalization of their counterparts in equilibrium thermodynamics.

\section{nonequilibrium partition function of Gaussian States}\label{E:gbejtejsd}

Here, our system is a harmonic oscillator with displacement $\chi$, which can represent the internal degrees of freedom of a harmonic atom or an Unruh-DeWitt detector in the dipole approximation. The external position $\mathbf{z}$ of this atom/detector is, for simplicity, assumed fixed in space. The bath which the atom/detector interacts with is modeled by a quantum massless scalar field $\phi$ initially prepared in its thermal state. Suppose that the system oscillator is initially in any arbitrary Gaussian state, thus out of equilibrium from the thermal bath, the subsequent evolution of the system oscillator is nonequilibrium in nature.

Many studies~\cite{GWT84,HWL08,CPR2n4D} have shown that when a system oscillator is bilinearly coupled to a linear thermal bath, it will eventually relax to an equilibrium state. In the weak coupling limit, this equilibrium state turns out to be the oscillator's thermal state. Throughout the nonequilibrium evolution, the reduced dynamics of the system oscillator is in general nonstationary but remains Gaussian. Thus in analogy with \eqref{E:dkbshfgbd}, to generalize the concept of the equilibrium partition function, we shall replace the density matrix operator of the thermal state of the system oscillator by that of a general Gaussian state.

The density matrix operator (not the matrix elements as shown in \eqref{E:nneker}) of a general Gaussian state takes the form~\cite{OL12,AA89,RW67,GA95,CD03}
\begin{equation}\label{E:bjdfhrse}
	\hat{\rho}_{\textsc{s}}=\hat{D}(\alpha)\hat{S}(\zeta)\hat{R}(\theta)\,\hat{\rho}_{\vartheta}\,\hat{R}^{\dagger}(\theta)\hat{S}^{\dagger}(\zeta)\hat{D}^{\dagger}(\alpha)\,,
\end{equation}
where  the subscripts $\mathrm{s}$ denotes the reduced system and $\vartheta$ emphasizes the system under nonequilibrium evolution. The operators $\hat{D}(\alpha)$, $\hat{S}(\zeta)$ and $\hat{R}(\theta)$ are respectively the displacement, squeeze and the rotation operators
\begin{align}
	\hat{D}(\alpha)&=\exp\biggl[\alpha\,\hat{a}^{\dagger}-\alpha^{*}\,\hat{a}\biggr]\,,&\hat{S}(\zeta)&=\exp\biggl[\frac{1}{2}\,\zeta^{*}\hat{a}^{2}-\frac{1}{2}\,\zeta\,\hat{a}^{\dagger\,2}\biggr]\,,&\hat{R}(\theta)&=\exp\biggl[-i\,\theta \Bigl(\hat{a}^{\dagger}\hat{a}+\frac{1}{2}\Bigr)\biggr]\,,
\end{align}
with the coherent parameter $\alpha$, the squeeze parameter $\zeta\in\mathbb{C}$ and the rotation angle $\theta\in\mathbb{R}$. The annihilation and creation operators $a$, $a^{\dagger}$ satisfies the canonical commutation relation $[\hat{a}, \hat{a}^{\dagger}]=1$ and are related to the position operator $\hat{\chi}$ and the momentum operator $\hat{p}$ of the oscillator by
\begin{align}\label{E:kaeawesf}
	\hat{\chi}&=\frac{1}{\sqrt{2m\omega_{\textsc{r}}}}\Bigl(\hat{a}^{\dagger}+\hat{a}\Bigr)\,,&\hat{p}&=i\sqrt{\frac{m\omega_{\textsc{r}}}{2}}\Bigl(\hat{a}^{\dagger}-\hat{a}\Bigr)\,,
\end{align}
where $m$, $\omega_{\textsc{r}}$ are the mass and the physical oscillating frequency of the oscillator. Here the operator $\hat{\rho}_{\vartheta}$ is a thermal-like state,
\begin{align}\label{E:bgkgere}
	\hat{\rho}_{\vartheta}&=\frac{1}{\mathcal{Z}_{\vartheta}}\,\exp\biggl[-\vartheta\,\Bigl(\hat{a}^{\dagger}\hat{a}+\frac{1}{2}\Bigr)\biggr]\,,&\mathcal{Z}_{\vartheta}&=\operatorname{Tr}\exp\biggl[-\vartheta\,\Bigl(\hat{a}^{\dagger}\hat{a}+\frac{1}{2}\Bigr)\biggr]=\frac{1}{2\sinh\frac{\vartheta}{2}}\,,
\end{align}
because at this stage the positive real parameter $\vartheta$ does not have the interpretation of the inverse temperature yet. The parameters $\alpha$, $\zeta$, $\theta$, and $\vartheta$ are completely arbitrary and thus can be in principle functions of time. A nice feature of \eqref{E:bjdfhrse} is that the trace
\begin{equation}
	\operatorname{Tr}\biggl\{\hat{D}(\alpha)\hat{S}(\zeta)\hat{R}(\theta)\,\exp\biggl[-\vartheta\,\Bigl(\hat{a}^{\dagger}\hat{a}+\frac{1}{2}\Bigr)\biggr]\,\hat{R}^{\dagger}(\theta)\hat{S}^{\dagger}(\zeta)\hat{D}^{\dagger}(\alpha)\biggr\}
\end{equation}
is independent of the included coherent, squeeze, and rotation operators, so it is still given by $\mathcal{Z}_{\vartheta}$. This suggests that $\mathcal{Z}_{\vartheta}$ will be our candidate of the nonequilibrium partition function, once we have identified \eqref{E:bjdfhrse} with the reduced density matrix of the oscillator system at an arbitrary moment and have expressed the parameters $\alpha$, $\zeta$, and $\vartheta$ in terms of the time-dependent covariance matrix elements. Note that $\theta$ in this case amounts to an arbitrary global phase of the state, so it will not enter in the following results.

Direct evaluation of \eqref{E:bjdfhrse} gives
\begin{align}
	&\quad\hat{\rho}_{\textsc{s}}(\hat{a},\hat{a}^{\dagger})\label{E:dkbruerw}\\
	&=\frac{2}{\sqrt{e^{2\varphi}-1}}\,\exp\biggl\{-2e^{-\varphi}\cosh^{-1}(\coth\varphi)\,\Bigl[\kappa\bigr(\hat{a}-\alpha\bigr)^{2}+\kappa^{*}\bigl(\hat{a}^{\dagger}-\alpha^{*}\bigr)^{2}+\frac{\lambda}{2}\,\bigl\{\hat{a}-\alpha,\,\hat{a}^{\dagger}-\alpha^{*}\bigr\}\Bigr]\biggr\}\,,\notag
\end{align}
where the squeeze parameter $\zeta$ is decomposed into $\zeta=\eta\,e^{i\,\psi}$ with $\eta\geq0$, $\psi\in\mathbb{R}$, and
\begin{align}\label{E:ekbddwwe}
	e^{2\varphi}&=\Xi^{2}\,,&\kappa&=\frac{\Xi}{4}\,\sinh2\eta\,e^{-i\psi}\,,&\lambda&=\frac{\Xi}{2}\,\cosh2\eta\,,\\
	\Xi&=\coth\frac{\vartheta}{2}\,,&\mathcal{Z}_{\vartheta}&=\frac{\sqrt{e^{2\varphi}-1}}{2}\,.
\end{align}
It will be identified later, from the matrix element of the density matrix, that the parameters $\lambda$, $\kappa$ in \eqref{E:dkbruerw} are related to the elements of the covariant matrix of the oscillator by
\begin{align}
	b&=\langle\hat{\chi}^{2}\rangle=2\bigl(\lambda-\kappa-\kappa^{*}\bigr)\mathfrak{b}^{2}=\frac{1}{2m\omega_{\textsc{r}}}\,\Xi\,\bigl(\cosh2\eta-\sinh2\eta\,\cos\psi\bigr)\,,\label{E:gfkj1}\\
	a&=\langle\hat{p}^{2}\rangle=2\bigl(\lambda+\kappa+\kappa^{*}\bigr)\mathfrak{a}^{2}=\frac{m\omega_{\textsc{r}}}{2}\,\Xi\,\bigl(\cosh2\eta+\sinh2\eta\,\cos\psi\bigr)\,,\\
	c&=\frac{1}{2}\langle\bigl\{\hat{\chi},\,\hat{p}\bigr\}\rangle=-2i\,\bigl(\kappa-\kappa^{*}\bigr)\mathfrak{a}\mathfrak{b}=-\frac{\Xi}{2}\,\sinh2\eta\,\sin\psi\,,\label{E:gfkj3}
\end{align}
with $e^{2\varphi}=4\bigl(\lambda^{2}-4\lvert\kappa\rvert^{2}\bigr)=4\bigl(ab-c^{2}\bigr)$ and
\begin{align}
	\mathfrak{a}&=\sqrt{\frac{m\omega_{\textsc{r}}}{2}}\,,&\mathfrak{b}&=\frac{1}{\sqrt{2m\omega_{\textsc{r}}}}\,,&\mathfrak{a}\mathfrak{b}&=\frac{1}{2}\,.
\end{align}
Since in our configuration the mean values of the canonical variables of the oscillator are zeros, we have set $\alpha=0$ such that the displacement operator does not play any role. Thus the reduced density operator describes a general \textit{squeezed thermal state}.

We then express the density matrix operator $\hat{\rho}_{\textsc{s}}$ in terms of the canonical operators $(\hat{\chi},\hat{p})$ of the oscillator by \eqref{E:kaeawesf} in order to facilitate subsequent evaluation of its matrix elements,
\begin{equation}\label{E:jterjbfdfs}
	\hat{\rho}_{\textsc{s}}(\hat{\chi},\hat{p})=\frac{2}{\sqrt{e^{2\varphi}-1}}\,\exp\biggl\{-e^{-\varphi}\cosh^{-1}(\coth\varphi)\,\Bigl[a\,\hat{\chi}^{2}+b\,\hat{p}^{2}-c\,\bigl\{\hat{\chi},\,\hat{p}\bigr\}\Bigr]\biggr\}\,.
\end{equation}
Note that the factor $e^{-\varphi}\cosh^{-1}\coth\varphi$ in \eqref{E:jterjbfdfs} can be related to the covariance matrix elements by
\begin{equation}\label{E:dgdjjwr}
	e^{-\varphi}\cosh^{-1}\coth\varphi=\frac{1}{2\sqrt{ab-c^{2}}}\,\ln\frac{\sqrt{ab-c^{2}}+\frac{1}{2}}{\sqrt{ab-c^{2}}-\frac{1}{2}}\,.
\end{equation}
The evaluation of the matrix elements of this density operator is carried out with the help of the McCoy theorem~\cite{MC32a,MC32b,HM94}
\begin{theorem}[McCoy theorem]
	Given the operators $\hat{\chi}$, $\hat{p}$ satisfying the commutation relation $[\hat{p},\hat{\chi}]=\mathfrak{c}$, and supposing that in the $\chi$-representation $\hat{p}$ will behave as a differential operator $\mathfrak{c}\,\partial/\partial\chi$, the operator function $\exp\Bigl(-A\,\hat{\chi}^{2}-B\,\hat{p}^{2}+2C\,\hat{\chi}\hat{p}\Bigr)$ in the standard ordering can be decomposed as
\begin{equation}\label{E:gbkdds}
	\exp\Bigl(-A\,\hat{\chi}^{2}-B\,\hat{p}^{2}+2C\,\hat{\chi}\hat{p}\Bigr)=\biggl(\frac{\varsigma}{\mathcal{A}\mathcal{D}\,e^{-2\mathfrak{c}\varsigma}-\mathcal{B}\mathcal{C}}\biggr)^{\frac{1}{2}}e^{-\mathcal{D}\mathfrak{c}}\,\exp\Bigl(\frac{Z}{2}\,\hat{\chi}^{2}\Bigr)\exp\Bigl(+Y\,\hat{\chi}\hat{p}\Bigr)_{\textsc{std}}\exp\Bigl(+\frac{X}{2}\,\hat{p}^{2}\Bigr)\,.
\end{equation}
with arbitrary parameters $A$, $B$, $C$, $\varsigma=\mathcal{A}\mathcal{D}-\mathcal{B}\mathcal{C}$,
\begin{align}
	X&=-\frac{1}{\mathfrak{c}}\frac{\mathcal{A}\mathcal{C}(e^{-2\mathfrak{c}\varsigma}-1)}{\mathcal{A}\mathcal{D}\,e^{-2\mathfrak{c}\varsigma}-\mathcal{B}\mathcal{C}}\,,&Y&=-\frac{1}{\mathfrak{c}}\biggl[1-\frac{\varsigma\,e^{-\mathfrak{c}\varsigma}}{\mathcal{A}\mathcal{D}\,e^{-2\mathfrak{c}\varsigma}-\mathcal{B}\mathcal{C}}\biggr]\,,&Z&=\frac{\mathcal{B}\mathcal{D}}{\mathcal{A}\mathcal{C}}\,X\,,
\end{align}
and
\begin{align*}
	\mathcal{A}&=1\,,&\mathcal{B}&=-\frac{1}{B}\Bigl(C-\sqrt{C^{2}-AB}\Bigr)\,,&\mathcal{C}&=-B\,,\\
	\mathcal{D}&=C+\sqrt{C^{2}-AB}\,,&\mathcal{E}&=-\mathcal{D}\,,&\mathcal{K}&=\biggl(\frac{\eta}{\mathcal{A}\mathcal{D}\,e^{-2\mathfrak{c}\eta}-\mathcal{B}\mathcal{C}}\biggr)^{\frac{1}{2}}\,.
\end{align*}
The subscript $\text{STD}$ refers to the standard ordering of the $\hat{\chi}$, $\hat{p}$ operators.
\end{theorem}
Then the matrix elements $\langle\chi\vert\,\exp\bigl(-A\,\hat{\chi}^{2}-B\,\hat{p}^{2}+2C\,\hat{\chi}\hat{p}\bigr)\,\vert\chi'\rangle$ are given by
\begin{align}\label{E:tuerfd}
	&\quad\langle\chi\vert\,\exp\Bigl(-A\,\hat{\chi}^{2}-B\,\hat{p}^{2}+2C\,\hat{\chi}\hat{p}\Bigr)\,\vert\chi'\rangle\notag\\
	&=\frac{\mathcal{K}}{\sqrt{-2\pi X}}\,e^{-\mathcal{D}\mathfrak{c}}\,\exp\Bigl[\frac{XZ-(Y+i)^{2}}{2X}\,\chi^{2}+\frac{i(Y+i)}{X}\,\chi\chi'+\frac{1}{2X}\,\chi'^{2}\Bigr]\,.
\end{align}
In the case of \eqref{E:jterjbfdfs}, we have chosen $\mathfrak{c}=-i$, and expressed the parameters in the McCoy theorem in terms of the covariance matrix elements
\begin{align*}
	A&=a\,e^{-\varphi}\cosh^{-1}\coth\varphi\,,&B&=b\,e^{-\varphi}\cosh^{-1}\coth\varphi\,,&C&=c\,e^{-\varphi}\cosh^{-1}\coth\varphi\,,
\end{align*}
with $a$, $b$, $c$ and $\varphi$ given by \eqref{E:ekbddwwe} and \eqref{E:gfkj1}--\eqref{E:gfkj3}. After some algebra,  we find
\begin{align}
	\mathcal{A}&=1\,,&\mathcal{B}&=\frac{-c+i\sqrt{ab-c^{2}}}{b}\,,\label{E:dgdjjwv1}\\
	\mathcal{C}&=-\frac{b}{2\sqrt{ab-c^{2}}}\,\ln\frac{\sqrt{ab-c^{2}}+\frac{1}{2}}{\sqrt{ab-c^{2}}-\frac{1}{2}}\,,&\mathcal{D}&=\frac{c+i\sqrt{ab-c^{2}}}{2\sqrt{ab-c^{2}}}\,\ln\frac{\sqrt{ab-c^{2}}+\frac{1}{2}}{\sqrt{ab-c^{2}}-\frac{1}{2}}\,,\\
	\eta&=i\,\ln\frac{\sqrt{ab-c^{2}}+\frac{1}{2}}{\sqrt{ab-c^{2}}-\frac{1}{2}}\,,&\mathcal{K}&=\frac{1+2\sqrt{ab-c^{2}}}{2\sqrt{ab-(c-i/2)^{2}}}\,,
\end{align}
such that
\begin{align}\label{E:dgdjjwv3}
	X&=-\frac{b}{ab-(c-\frac{i}{2})^{2}}\,,&Y&=\frac{c-\frac{i}{2}}{ab-(c-\frac{i}{2})^{2}}\,,&Z&=-\frac{a}{ab-(c-\frac{i}{2})^{2}}\,.
\end{align}
From \eqref{E:tuerfd}, we find the elements of the density matrix operator \eqref{E:jterjbfdfs} take on the form
\begin{align}
	\langle\chi\vert\,\,\hat{\rho}_{\textsc{s}}(\hat{\chi},\hat{p})\,\vert \chi'\rangle&=\frac{1}{\sqrt{2\pi b}}\,\exp\biggl[-\frac{ab-(c+\frac{i}{2})^{2}}{2b}\,\chi^{2}+\frac{ab-c^{2}-\frac{1}{4}}{b}\,\chi\chi'-\frac{ab-(c-\frac{i}{2})^{2}}{2b}\,\chi'^{2}\biggr]\,,\label{E:eurbsdwe}
\end{align}
after we plug in the parameters in \eqref{E:dgdjjwv1}--\eqref{E:dgdjjwv3} with $b=\langle\hat{\chi}^{2}\rangle$, $a=\langle\hat{p}^{2}\rangle$, and $c=\dfrac{1}{2}\langle\bigl\{\hat{\chi},\,\hat{p}\bigr\}\rangle$. Eq.~\eqref{E:eurbsdwe} can be further simplified if we use the CoM ($\Sigma$) and relative ($\Delta$) variables defined by $\chi=\Sigma+\Delta/2$ and $\chi'=\Sigma-\Delta/2$, respectively,
\begin{equation}
	\rho_{\textsc{s}}(\chi,\chi')=\langle\chi\vert\,\hat{\rho}_{\textsc{s}}(\hat{\chi},\hat{p})\,\vert \chi'\rangle=\frac{1}{\sqrt{2\pi b}}\,\exp\biggl[-\frac{1}{2b}\,\Sigma^{2}+i\,\frac{c}{b}\,\Sigma\Delta-\frac{ab-c^{2}}{2b}\,\Delta^{2}\biggr]\,.
\end{equation}
Thus we recover the familiar result for the Gaussian state in \eqref{E:nneker} for the case $\langle\hat{\chi}\rangle=0=\langle\hat{p}\rangle$.

The previous results allow us to uniquely relate the parameter $\vartheta$ and the squeeze parameter $\zeta$ with the covariance matrix elements associated with $\hat{\rho}_{\textsc{s}}$. Thus the nonequilibrium partition function $\mathcal{Z}_{\textsc{s}}$ of the reduced system in \eqref{E:bgkgere} can be written as
\begin{align}\label{E:gjhdfsjwe}
	\mathcal{Z}_{\textsc{s}}=\mathcal{Z}_{\vartheta}&=\frac{\sqrt{e^{2\varphi}-1}}{2}=\Bigl(ab-c^{2}-\frac{1}{4}\Bigr)^{\frac{1}{2}}=\frac{1}{2}\,\operatorname{csch}\frac{\vartheta}{2}\,,
\end{align}
again, with $b=\langle\hat{\chi}^{2}\rangle$, $a=\langle\hat{p}^{2}\rangle$, and $c=\dfrac{1}{2}\langle\bigl\{\hat{\chi},\,\hat{p}\bigr\}\rangle$. Clearly it looks similar to \eqref{E:fkgbrtss}, and the expression inside the braces is connected to the \textit{Robertson-Schr\"odinger uncertainty relation}
\begin{equation}
	\langle\hat{\chi}^{2}\rangle\langle\hat{p}^{2}\rangle-\dfrac{1}{4}\langle\bigl\{\hat{\chi},\,\hat{p}\bigr\}\rangle^{2}\geq\frac{1}{4}\,.
\end{equation}
The uncertainty principle ensures the reality of the nonequilibrium partition function in the close-time path/in-in formalism.

We observe that (1) Eq.~\eqref{E:gjhdfsjwe} is an immediate generalization of \eqref{E:nfgjits} by including the correlation between the canonical variables of the system, (2) it is manifestly time-dependent because for the nonequilibrium evolution of the system coupled to the thermal bath, the covariance matrix elements of the system change with time, and (3) it points out a profound connection between  quantum non-commutativity and statistical mechanics. Violation of this uncertainty relation engenders unphysical results for the nonequilibrium partition function, and in turn the nonequilibrium free energy, which is proportional to the logarithm of the nonequilibrium partition function. Any thermodynamic function derived from them, such as entropy, flawed, thus endangering the consistency of the thermodynamic relations. Finally we note that the calculations we have performed make no approximations and the results we obtained are exact, so they are valid for strong coupling between the system and the bath.

\section{Nonequilibrium Effective Temperature}	

\begin{figure}
\centering
    \scalebox{0.35}{\includegraphics{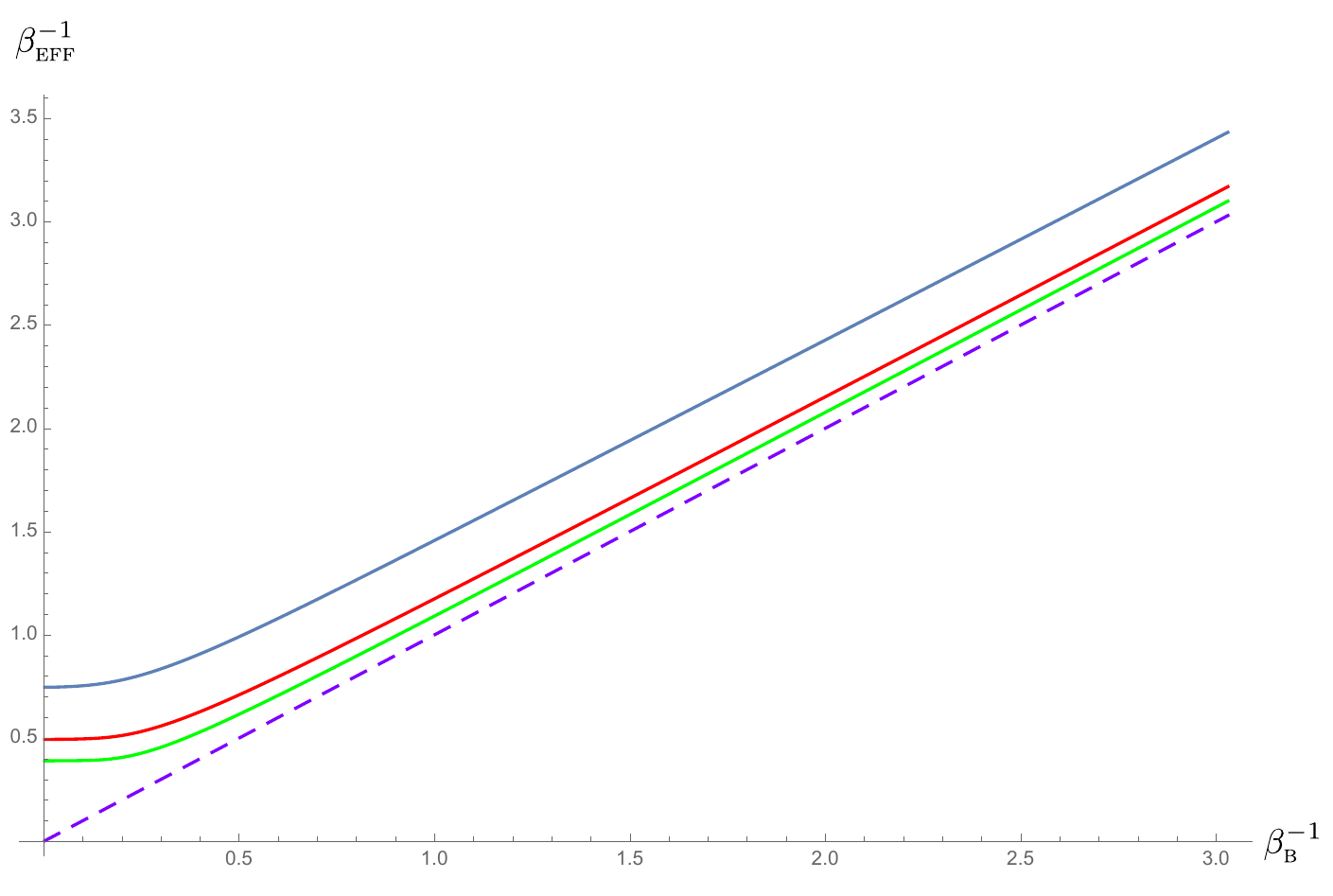}}
    \caption{The dependence of the effective temperature $\beta_{\textsc{eff}}^{-1}$ on the initial temperature $\beta_{\textsc{b}}^{-1}$ of the bath. The parameters are normalized with respect to the resonance frequency $\Omega=\sqrt{\omega_{\textsc{r}}^{2}-\gamma^{2}}$ such that $m=1\,\Omega$, $\gamma\,t=3.6$, and the cutoff frequency $\Lambda=1000\,\Omega$. The parameter $\gamma=e^{2}/8\pi m$ is the damping constant. We choose the initial displacement dispersion $\langle\hat{\chi}^{2}(0)\rangle=1/2$ and the momentum dispersion $\langle\hat{p}^{2}(0)\rangle=1/2$. We choose $\gamma=0.3$ for the blue solid curve, $\gamma=0.1$ for the orange solid curve, and $\gamma=0.03$ for the green solid curve. The evolution time is long enough such that the dynamics is sufficiently relaxed. The dashed line represents the reference to the initial temperature of the bath.}\label{Fi:effectiveT}
\end{figure}

We see that the nonequilibrium partition function $\mathcal{Z}_{\textsc{s}}$, and the covariance matrix elements of the oscillator system at any time can be related to the temperature-like parameter $\vartheta$ in a way analogous to their counterparts in Sec.~\ref{S:ohtbfgd} when the system stays in an equilibrium thermal state. Thus we can introduce the effective temperature $\beta_{\textsc{eff}}^{-1}(t)=\omega_{\textsc{r}}/\vartheta(t)$, where $\omega_{\textsc{r}}$ is the physical frequency of the system oscillator. A question remains, whether the nonequilibrium partition function and the covariance matrix elements may introduce inequivalent effective temperatures. We observe that from the analogy between the general Gaussian state and the thermal state, it is not sufficient to identify the effective temperature by the covariance matrix elements $\langle\hat{\chi}^{2}\rangle$ and $\langle\hat{p}^{2}\rangle$ alone, because $\langle\{\hat{\chi},\hat{p}\}\rangle$ vanishes in the thermal state but not in the general Gaussian state. Thus we will use the nonequilibrium partition function, such that the effective temperature $\beta_{\textsc{eff}}^{-1}(t)$ satisfies
\begin{align}\label{E:fgrhjss}
	2\sinh\frac{\beta_{\textsc{eff}}\omega_{\textsc{r}}}{2}=\Bigl(ab-c^{2}-\frac{1}{4}\Bigr)^{-\frac{1}{2}}\,.
\end{align}
Note that since the nonequilibrium partition function defined this way is always real and non-negative, the corresponding effective temperature $\beta_{\textsc{eff}}^{-1}(t)$ possesses the same property for the Gaussian system. The effective temperature introduced here has a \textit{dynamic} significance more so than a statistical one. From its construction, we observe that even before the interaction is turned on, we can assign an effective temperature to the initial state of the system. It may not have any statistical meaning as that in an equilibrium thermal state at this stage. When we turn on the interaction between the oscillator and a thermal bath of initial temperature $\beta_{\textsc{b}}^{-1}$, the effective temperature starts evolving and moves toward the initial value of the bath temperature after the oscillator gradually settles down to an equilibrium state. In general it will approach a constant different from the initial temperature of the bath, unless the oscillator-bath coupling is vanishingly weak. At this final asymptotic state of the oscillator, the effective temperature gradually loses the characteristics of the oscillator's initial state, but assumes the statistical nature of the thermal bath, skewed by the finite oscillator-bath coupling. Not until this regime will this time-dependent parameter $\beta_{\textsc{eff}}^{-1}$ acquire the true meaning of temperature defined in equilibrium thermodynamics. This is particularly clearly seen in the weak oscillator-bath coupling limit, shown in Fig.~\ref{Fi:effectiveT}. There, the dependence of the effective temperature on the initial temperature $\beta_{\textsc{b}}^{-1}$ of the bath is drawn for three different oscillator-bath coupling strengths. The weaker the coupling is, the closer the solid curve approaches to the dashed straight line, representing  the  initial temperature of the bath. In addition, in the high bath temperature limit, all three curves have the same slope unity, signaling that the quantum effects become subdominant. In contrast, in the low bath temperature limit, the curves significantly deviate from the dashed line. This is the consequence of finite system-bath coupling and is related to the quantum entanglement between the oscillator system and the field bath. In particular, the effective temperature is not zero even when the oscillator is initially coupled to a zero-temperature bath. This point will become clearer once we derive the von Neumann entropy of the oscillator system~\cite{QTD2}. It also has some interesting implications in the interpretation of the third law with entanglement considerations. .

Even though the meaning of the effective temperature is rather obscure during the nonequilibrium evolution, it can be used to provide a unified formulation of nonequilibrium quantum thermodynamics on a par with the equilibrium thermodynamics we have been familiar with. In formulating the nonequilibrium thermodynamics at finite coupling, numerous ambiguities arise in the choices of thermodynamic potentials and temperatures of the system~\cite{QTDpot,MU19,TH20}. Consistency may be inadvertently lost within a myriad of their combinations. For example, in the expression $\mathcal{F}=\mathcal{U}-T\,\mathcal{S}$, there are more than one possibility for the choice of the internal energy $\mathcal{U}$, the temperature $T$ and the entropy $\mathcal{S}$, given that the free energy $\mathcal{F}$ is known. Thus  a more detailed investigation is needed into what qualifies as a sensible combination.

The introduction of the effective temperature allows us to identify a Hamiltonian of the mean force~\cite{HTL11,HTL11a} in the exponent in \eqref{E:jterjbfdfs} during the nonequilibrium evolution of the system
\begin{equation}\label{E:bgberjs}
	\hat{H}_{\textsc{mf}}=\beta_{\textsc{eff}}^{-1}\biggl\{e^{-\varphi}\cosh^{-1}(\coth\varphi)\,\Bigl[a\,\hat{\chi}^{2}+b\,\hat{p}^{2}-c\,\bigl\{\hat{\chi},\,\hat{p}\bigr\}\Bigr]\biggr\}\,,
\end{equation}
so that the density matrix \eqref{E:jterjbfdfs} take on a familiar form
\begin{equation}\label{E:flsjerdf}
	\hat{\rho}_{\textsc{s}}(t)=\frac{1}{\mathcal{Z}_{\textsc{s}}(t)}\,\exp\Bigl[-\beta_{\textsc{eff}}(t)\hat{H}_{\textsc{mf}}(t)\Bigr]\,,
\end{equation}
compared to the equilibrium thermal state, except for that here in \eqref{E:flsjerdf} every quantity is time-dependent. Note that the factor $e^{-\varphi}\cosh^{-1}\coth\varphi$ in \eqref{E:bgberjs} can be related to the covariance matrix elements by \eqref{E:dgdjjwv1}. We can do the same thing for the initial state of the oscillator. Such defined Hamiltonian of the mean force bears no resemblance to the Hamiltonian of the isolated oscillator. It and the effective temperature at this stage are more like paraphrasing the Gaussian state in a form mimicking the equilibrium thermal state. The similar modus operandi is widely used. We want to emphasize that \eqref{E:bgberjs} is by no means stationary even though it looks like an equilibrium thermal state. Accordingly, the expectation value $\langle\bigl\{\hat{\chi},\,\hat{p}\bigr\}\rangle\neq0$.

Once we identify the effective temperature, we can then introduce the nonequilibrium free energy $\mathcal{F}_{\textsc{s}}$ by
\begin{equation}\label{E:rbghdsfg}
	\mathcal{F}_{\textsc{s}}(t)=-\beta_{\textsc{eff}}^{-1}(t)\ln\mathcal{Z}_{\textsc{s}}(t)=\frac{\omega_{\textsc{r}}}{2}+\frac{1}{\beta_{\textsc{eff}}(t)}\,\ln\bigl(1-e^{-\beta_{\textsc{eff}}(t)\omega_{\textsc{r}}}\bigr)\,.
\end{equation}
This is an analogy of \eqref{E:bgruwia}, and will be extremely essential for the development of nonequilibrium quantum thermodynamics.

The covariance matrix elements $a$, $b$ and $c$ can be quite readily found by the Langevin equation for the reduced dynamics of the oscillator system~\cite{HWL08}. For example,
\begin{align*}
	b(t)&=\langle\hat{\chi}^{2}(t)\rangle\notag\\
	&=d_{1}^{2}(t)\,\langle\hat{\chi}^{2}(0)\rangle+d_{2}^{2}(t)\,\langle\hat{p}^{2}(0)\rangle+\frac{e^{2}}{m^{2}}\int_{0}^{t}\!ds\,ds'\;d_{2}(t-s)d_{2}(t-s')\,G_{H}^{(\phi)}(\mathbf{z},s;\mathbf{z},s')\,,\\
	a(t)&=\langle\hat{p}^{2}(t)\rangle\\
	&=m^{2}\,\dot{d}_{1}^{2}(t)\,\langle\hat{\chi}^{2}(0)\rangle+m^{2}\,\dot{d}_{2}^{2}(t)\,\langle\hat{p}^{2}(0)\rangle+e^{2}\int_{0}^{t}\!ds\,ds'\;\dot{d}_{2}(t-s)\dot{d}_{2}(t-s')\,G_{H}^{(\phi)}(\mathbf{z},s;\mathbf{z},s')\,,\\
	c(t)&=\frac{1}{2}\langle\bigl\{\hat{\chi}(t),\,\hat{p}(t)\bigr\}\rangle\\
	&=m\,d_{1}(t)\dot{d}_{1}(t)\,\langle\hat{\chi}^{2}(0)\rangle+m\,d_{2}(t)\dot{d}_{2}(t)\,\langle\hat{p}^{2}(0)\rangle+\frac{e^{2}}{m}\int_{0}^{t}\!ds\,ds'\;d_{2}(t-s)\dot{d}_{2}(t-s')\,G_{H}^{(\phi)}(\mathbf{z},s;\mathbf{z},s')\,,
\end{align*}
where for simplicity we have assumed that $\hat{\chi}$ and $\hat{p}$ are not initially correlated and that $\langle\hat{\chi}(0)\rangle=0=\langle\hat{p}(0)\rangle$. The two fundamental homogeneous solutions $d_{1}(t)$ and $d_{2}(t)$ of the Langevin equation
\begin{equation}\label{E:gbjekrd}
	\ddot{\hat{\chi}}(t)+2\gamma\,\dot{\hat{\chi}}(t)+\omega_{\textsc{r}}^{2}\hat{\chi}(t)=\frac{e}{m}\,\hat{\phi}_{h}(\mathbf{z},t)\,,
\end{equation}
obey the initial conditions
\begin{align}
	d_{1}(0)&=1\,,&\dot{d}_{1}(0)&=0\,,&d_{2}(0)&=0\,,&\dot{d}_{2}(0)&=1\,,
\end{align}
at the initial time $t=0$. Eq.~\eqref{E:gbjekrd} describes the reduced dynamics of the internal degree of freedom of an atom/Unruh-DeWitt detector at the spatial location $\mathbf{z}$, when it bilinearly couples with the a free massless quantum  scalar field $\hat{\phi}_{h}$ with the coupling strength $e$. As mentioned earlier, we model the internal degree of freedom by the harmonic oscillator of mass $m$ and physical oscillating frequency $\omega_{\textsc{r}}$. The damping constant $\gamma=e^{2}/8\pi m$ signifies the relaxation time scale and the coupling strength. The associated quantum frictional force $-2m\gamma\,\dot{\hat{\chi}}(t)$ and the quantum noise force $e\,\hat{\phi}_{h}(\mathbf{z},t)$ results from the interaction between the detector and the bath field. This dissipative backaction will be self-consistently paired with the quantum noise force in a definite way~\cite{CPR2n4D,HWL08}. Since we assume that the bath field is initially in its thermal state, described the density matrix $\hat{\varrho}_{\beta}^{(\phi)}$ of temperature $\beta_{\textsc{b}}^{-1}$, the free field operator $\hat{\phi}_{h}(\mathbf{z},t)$ has these properties
\begin{align}
	\langle\hat{\phi}_{h}(\mathbf{z},t)\rangle&=0\,,\\
	\frac{1}{2}\,\langle\bigl\{\hat{\phi}_{h}(\mathbf{z},t),\hat{\phi}_{h}(\mathbf{z},t)\bigr\}\rangle&=\frac{1}{2}\,\operatorname{Tr}_{\phi}\Bigl[\hat{\varrho}_{\beta}^{(\phi)}\bigl\{\hat{\phi}_{h}(\mathbf{z},t),\hat{\phi}_{h}(\mathbf{z},t')\bigr\}\Bigr]=G_{H}^{(\phi)}(\mathbf{z},t;\mathbf{z},t')\,.\label{E:gbsecvs}
\end{align}
In \eqref{E:gbsecvs}, the trace is carried out over the field degree of freedom and all the higher moments of the free field can be expanded by these first two moments. This and the assumption about the initial state of the system explain why we do not need to consider the displacement operator. It will not pose any problem to our discussions but greatly simplify the mathematical presentations.

\section{Nonequilibrium free energy and influence action}\label{S:gbseruhds}
 
Now we turn to the final item in our wish-list. That is, can we use the close-time path integral formalism to generalize \eqref{E:dkbshfgbd} and assimilate nonequilibrium free energy within the formal structure of nonequilibrium quantum field theory? Observing from \eqref{E:bjdfhrse} and \eqref{E:jterjbfdfs}, we find
\begin{align}
	\mathcal{Z}_{\textsc{s}}&=\operatorname{Tr}_{\chi}\biggl\{\hat{D}(\alpha)\hat{S}(\sigma)\hat{R}(\theta)\,\exp\biggl[-\vartheta\,\Bigl(\hat{a}^{\dagger}\hat{a}+\frac{1}{2}\Bigr)\biggr]\,\hat{R}^{\dagger}(\theta)\hat{S}^{\dagger}(\sigma)\hat{D}^{\dagger}(\alpha)\biggr\}\notag\\
	&=\operatorname{Tr}_{\chi}\exp\biggl[-A\,\hat{\chi}^{2}-B\,\hat{p}^{2}+C\,\bigl\{\hat{\chi},\,\hat{p}\bigr\}\biggr]\,,\label{E:jjfgbd}
\end{align}
for a general Gaussian state, since $\operatorname{Tr}_{\chi}\hat{\rho}_{\textsc{s}}=1$, with
\begin{align*}
	A&=a\,e^{-\varphi}\cosh^{-1}(\coth\varphi)\,,&B&=b\,e^{-\varphi}\cosh^{-1}(\coth\varphi)\,,&C&=c\,e^{-\varphi}\cosh^{-1}(\coth\varphi)\,\,,
\end{align*}
and the covariance matrix elements $a$, $b$, $c$, $\varphi$ are related to the squeeze parameter $\zeta$, and nonequilibrium parameter $\vartheta$ by \eqref{E:ekbddwwe}--\eqref{E:gfkj1}. This establishes the first equal sign in \eqref{E:dkbshfgbd} for the nonequilibrium Gaussian system.

To establish the second equality in \eqref{E:dkbshfgbd} for the nonequilibrium Gaussian system, we note that in the open system framework the reduced density matrix elements \eqref{E:eurbsdwe} can be represented by the forward- and backward-time-path integrals as
\begin{align}\label{E:bdkfbs}
	\rho_{\textsc{s}}(\chi,\chi')=\int\!d\chi_{i}\,d\chi'_{i}\int_{\chi_{i}}^{\chi}\!\mathcal{D}\chi_{+}\int_{\chi'_{i}}^{\chi'}\!\mathcal{D}\chi_{-}\;\exp\biggl\{i\,S_{\chi}[\chi_{+}]-i\,S_{\chi}[\chi_{-}]+i\,S_{\textsc{if}}[\chi_{+},\chi_{-}]\biggr\}\,\rho_{\textsc{s}}(\chi_{i}^{\vphantom{'}},\chi'_{i})\,.
\end{align}
The functions $\chi_{\pm}(t)$ denote $\chi$ evaluated at the forward (backward) real-time path, the variables $\chi^{\vphantom{'}}_{i}$, $\chi'_{i}$ represent different initial values of $\chi$, and $S_{\chi}$ is the (real-time, not Euclidean) action of the free system in the absence of interaction with the bath. The influence action $S_{\textsc{if}}[\chi_{+},\chi_{-}]$ in the current case for a Gaussian open system takes the form
\begin{align}
	S_{\textsc{if}}[\chi_{+},\chi_{-}]&=e^{2}\int_{0}^{t}\!ds\int_{0}^{t}\!ds'\;\Delta^{(\chi)}(s)\,G_{R}^{(\phi)}(s,s')\,\tau^{(\chi)}(s')\notag\\
	&\qquad\qquad\qquad\qquad\qquad+i\,\frac{e^{2}}{2}\int_{0}^{t}\!ds\int_{0}^{t}\!ds'\;\Delta^{(\chi)}(s)\,G_{H}^{(\phi)}(s,s')\,\Delta^{(\chi)}(s')\,,
\end{align}
where $\Delta^{(\chi)}=\chi_{+}-\chi_{-}$, $\Sigma^{(\chi)}=(\chi_{+}+\chi_{-})/2$, and $G_{R}^{(\phi)}(t,t')$ is the retarded Green's function, or dissipation kernel, of the free scalar field $\hat{\phi}_{h}$,
\begin{equation}
	G_{R}^{(\phi)}(t,t')=G_{R}^{(\phi)}(\mathbf{z},t;\mathbf{z},t')= i\,\theta(t-t')\,\bigl[\hat{\phi}_{h}(\mathbf{z},t),\hat{\phi}_{h}(\mathbf{z},t')\bigr]\,,
\end{equation}
a $c$-number function independent of the initial state of the field. A similar shorthand notation applies to $G_{H}^{(\phi)}(t,t')$. Essentially it summarizes the effects of the bath field on the system of interest in a self-consistent way. It can be understood as the close-time-path effective action of the bath field driven by the coupling with the system
\begin{align}
	e^{i\,S_{\textsc{if}}[\chi_{+},\chi_{-}]}&=\int_{-\infty}^{\infty}\!d\phi\int_{-\infty}^{\infty}\!d\phi^{\vphantom{'}}_{i}\,d\phi'_{i}\;\rho_{\phi}(\phi^{\vphantom{'}}_{i},\phi'_{i})\int_{\phi_{i}}^{\phi}\!\mathcal{D}\phi_{+}\int_{\phi'_{i}}^{\phi}\!\mathcal{D}\phi_{-}\notag\\
	&\qquad\qquad\qquad\times\exp\biggl\{i\,S_{\phi}[\phi_{+}]-i\,S_{\phi}[\phi_{-}]+i\,S_{\textsc{i}}[\phi_{+},\chi_{+}]-i\,S_{\textsc{i}}[\phi_{-},\chi_{-}]\biggr\}\notag\\
	&=\int_{-\infty}^{\infty}\!d\phi^{\vphantom{'}}_{i}\,d\phi'_{i}\oint_{\phi_{i}}^{\phi'_{i}}\!\mathcal{D}\phi\;\exp\biggl\{i\,S_{\phi}[\phi]+i\,S_{\textsc{i}}[\phi,\chi]\biggr\}\,\rho_{\phi}(\phi_{i}^{\vphantom{'}},\phi'_{i})\,.\label{E:dgbkdjer}
\end{align}
We form a loop path integral by concatenating the forward and backward time paths. Here $S_{\textsc{i}}[\phi,\chi]$ is the bilinear interaction term between the bath and the system. Now the coarse-grained effective action $S_{\textsc{cg}}[\chi_{+},\chi_{-}]$ for the system is defined by~\cite{HPZ,JH1,CalHu08},
\begin{equation}
	S_{\textsc{cg}}[\chi_{+},\chi_{-}]=S_{\chi}[\chi_{+}]-S_{\chi}[\chi_{-}]+S_{\textsc{if}}[\chi_{+},\chi_{-}]\,.
\end{equation}
Then the trace of \eqref{E:bdkfbs} takes a compact form
\begin{align}
	&\quad\int_{-\infty}^{\infty}\!d\chi\,d\chi'\;\delta(\chi-\chi')\,\rho_{\textsc{s}}(\chi,\chi')=1\notag\\
	&=\int_{-\infty}^{\infty}\!d\chi\int\!d\chi^{\vphantom{'}}_{i}\,d\chi'_{i}\int_{\chi_{i}}^{\chi}\!\mathcal{D}\chi_{+}\int_{\chi'_{i}}^{\chi}\!\mathcal{D}\chi_{-}\;\exp\biggl\{i\,S_{\textsc{CG}}[\chi_{+},\chi_{-}]\biggr\}\,\rho_{\textsc{s}}(\chi_{i}^{\vphantom{'}},\chi'_{i})\,.
\end{align}
In analogy with \eqref{E:dgbkdjer}, this implies that the above path integrals give a result proportional to some sort of close-time-path effective action.

Since we have the operator equation
\begin{equation}
	\exp\biggl[-A\,\hat{\chi}^{2}-B\,\hat{p}^{2}+C\,\bigl\{\hat{\chi},\,\hat{p}\bigr\}\biggr]=\mathcal{Z}_{\textsc{s}}\,\hat{\rho}_{\textsc{s}}
\end{equation}
from \eqref{E:jterjbfdfs}, we thus formally arrive at
\begin{align*}
	\mathcal{Z}_{\textsc{s}}\,\rho_{\textsc{s}}(\chi,\chi')&=\langle\chi\vert\,\exp\biggl[-A\,\hat{\chi}^{2}-B\,\hat{p}^{2}+C\,\bigl\{\hat{\chi},\,\hat{p}\bigr\}\biggr]\,\vert\chi'\rangle\notag\\
	&=\mathcal{Z}_{\textsc{s}}\int\!d\chi_{i}\,d\chi'_{i}\int_{\chi_{i}}^{\chi}\!\mathcal{D}\chi_{+}\int_{\chi'_{i}}^{\chi'}\!\mathcal{D}\chi_{-}\;\exp\biggl\{i\,S_{\textsc{CG}}[\chi_{+},\chi_{-}]\biggr\}\,\rho_{\textsc{s}}(\chi_{i}^{\vphantom{'}},\chi'_{i})\,,
\end{align*}
such that after taking the trace with respect to $\chi$, Eq.~\eqref{E:jjfgbd} implies
\begin{equation}\label{E:djghs}
	\mathcal{Z}_{\textsc{s}}=\mathcal{N}(t)\int\!d\chi^{\vphantom{'}}_{i}\,d\chi'_{i}\oint_{\chi_{i}}^{\chi'_{i}}\!\mathcal{D}\chi\;\exp\biggl\{i\,S_{\textsc{CG}}[\chi_{+},\chi_{-}]\biggr\}\,\rho_{\textsc{s}}(\chi_{i}^{\vphantom{'}},\chi'_{i})\,,
\end{equation}
where the time path runs from $t=0$ to $t$ and then returns to $t=0$ again, a consequence of $\operatorname{Tr}_{\chi}\hat{\rho}_{\textsc{s}}(t)=1$. This is a formal generalization of \eqref{E:dkbshfgbd} from the imaginary-time path integral formalism to the close-time path integral formalism, and from equilibrium dynamics to nonequilibrium dynamics. The normalization function $\mathcal{N}(t)$ is such that the lefthand side of \eqref{E:djghs} gives the nonequilibrium partition function $\mathcal{Z}_{\textsc{s}}$. {This may sound trivial because in practice the normalization factor $\mathcal{N}(t)$ still has to be determined by the method discussed in Sec.~\ref{E:gbejtejsd}. However, this expression provides a formal intermediary from which we can generalize  previous results for   nonlinear systems by the perturbative functional method, such as outlined in~\cite{HHNL20a,HHNL20b}}.

It is interesting to note that if we also write the initial general Gaussian state $\hat{\rho}_{\textsc{s}}(0)$ in \eqref{E:djghs} in a form similar to \eqref{E:flsjerdf}
\begin{equation}
	\hat{\rho}_{\textsc{s}}(0)=\frac{1}{\mathcal{Z}_{\textsc{s}}(0)}\,\exp\Bigl[-\beta_{\textsc{eff}}(0)\hat{H}_{\textsc{mf}}(0)\Bigr]\,,
\end{equation}
where $\hat{H}_{\textsc{mf}}$ is the Hamiltonian of mean force defined in \eqref{E:bgberjs}, then \eqref{E:djghs} can be written as
\begin{align}\label{E:gfbkfjdfd}
	&\quad\beta_{\textsc{eff}}(t)\mathcal{F}_{\textsc{s}}(t)-\beta_{\textsc{eff}}(0)\mathcal{F}_{\textsc{s}}(0)\notag\\
	&=\ln\biggl[\frac{1}{\mathcal{N}(t)}\int\!d\chi_{i}\,d\chi'_{i}\oint_{\chi_{i}}^{\chi'_{i}}\!\mathcal{D}\chi\;\exp\biggl\{i\,S_{\textsc{CG}}[\chi_{+},\chi_{-}]\biggr\}\,\langle\chi_{i}\vert\,e^{-\beta_{\textsc{eff}}(0)\hat{H}_{\textsc{mf}}(0)}\,\vert\chi'_{i}\rangle\biggr]\,.
\end{align}
This is the ``difference'' of the nonequilibrium free energy of the reduced system with respect to its initial state. This may be linked to the fluctuation theorem in the case where there is no external drive, but viewed as a generalization when the final state is not an equilibrium state. This is because the close-time-path integral include the combined effects of all possible forward and backward trajectories connecting the initial and the final configurations. We note that the matrix elements in the integrand in~\eqref{E:gfbkfjdfd} is not an exponential form, as shown in~\eqref{E:eurbsdwe}. That is, in general the matrix element of the an exponential operator is not of an exponential function per se; it will yield an additional normalization factor, and that makes the determination of the normalization in~\eqref{E:djghs} less trivial than it appears. Finally, in contrast to the classical case where only the diagonal elements (measuring the probabilities) exist, for quantum systems, the off-diagonal elements of the density matrix also contribute to the expression involving time evolution.

From this effective free energy, we can introduce the corresponding nonequilibrium thermodynamic entropy for the reduced system,
\begin{equation}
	\mathcal{S}_{\textsc{s}}=\beta^{2}_{\textsc{eff}}\frac{\partial\mathcal{F}_{\textsc{s}}}{\partial\beta_{\textsc{eff}}}\,,
\end{equation}
which can be shown~\cite{QTD2} to  coincide with the von Neumann entropy associated the reduced density matrix by
\begin{equation}
	\mathcal{S}_{vN}=-\operatorname{Tr}_{\chi}\Bigl\{\hat{\rho}_{\textsc{s}}\,\ln\hat{\rho}_{\textsc{s}}\Bigr\}\,.
\end{equation}
This enables us to introduce the internal energy $\mathcal{U}_{\textsc{s}}$ by $\mathcal{F}_{\textsc{s}}=\mathcal{U}_{\textsc{s}}-\beta^{-1}_{\textsc{eff}}\mathcal{S}_{\textsc{s}}$
\begin{equation}\label{E:gndbsds}
	\mathcal{U}_{\textsc{s}}=\mathcal{F}_{\textsc{s}}+\beta^{-1}_{\textsc{eff}}\mathcal{S}_{\textsc{s}}=\frac{\omega_{\textsc{r}}}{2}\,\coth\frac{\beta_{\textsc{eff}}\omega_{\textsc{r}}}{2}\,.
\end{equation}
This expression is of the same form as the internal energy of an oscillator in its thermal state in the conventional (weak coupling) thermodynamics except that the oscillator's temperature is replaced by the  nonequilibrium effective temperature $\beta_{\textsc{eff}}^{-1}$. Notably it satisfies
\begin{equation}
	\mathcal{U}_{\textsc{s}}=-\frac{\partial}{\partial\beta_{\textsc{eff}}}\ln\mathcal{Z}_{\textsc{s}}\,,\label{E:rubsgbfrt}
\end{equation}
and we can straightforwardly show that
\begin{equation}
	\biggl(\frac{\partial\mathcal{U}}{\partial\mathcal{S}}\biggr)=\beta_{\textsc{eff}}^{-1}\,,
\end{equation}
so that the effective temperature satisfies the conventional thermodynamic relation derived from the internal energy and the entropy. This is expected formally since our nonequilibrium effective temperature is defined via the nonequilibrium partition function. Note that all these aforementioned quantities are time-dependent and the relations hold at any moment in the nonequilibrium evolution, not just restricted to the equilibrium condition as in conventional thermodynamics.

The internal energy can be shown to be the quantum expectation value of the Hamiltonian of mean force \eqref{E:bgberjs}
\begin{equation}
	\mathcal{U}_{\textsc{s}}=\langle\hat{H}_{\textsc{mf}}\rangle=\operatorname{Tr}_{\chi}\Bigl\{\hat{\rho}_{\textsc{s}}\,\hat{H}_{\textsc{mf}}\Bigr\}\,.
\end{equation}
However, it is not equal to the expectation value of the system's Hamiltonian operators~\cite{HMF}. From the representation of the reduced density matrix \eqref{E:bjdfhrse}, we can see the connection between the two Hamiltonian operators $\hat{H}_{\textsc{mf}}$ and $\hat{H}_{\textsc{s}}$
\begin{align}
	e^{-\beta_{\textsc{eff}}\hat{H}_{\textsc{mf}}}&=\hat{S}(\zeta)\,e^{-\beta_{\textsc{eff}}\hat{H}_{\textsc{s}}}\,\hat{S}^{\dagger}(\zeta)=e^{-\beta_{\textsc{eff}}\,\hat{S}(\zeta)\hat{H}_{\textsc{s}}\hat{S}^{\dagger}(\zeta)}\,,&&\Rightarrow&\hat{H}_{\textsc{mf}}&=\hat{S}(\zeta)\hat{H}_{\textsc{s}}\hat{S}^{\dagger}(\zeta)\,.
\end{align}
This shows how interaction with the bath enters in the Hamiltonian of mean force via the squeeze parameter.

Here it is interesting to compare our new findings with the earlier results on thermodynamics at strong coupling~\cite{GT09,Se16,JA17,QTDpot}. In theories assuming the entire system is in a global thermal state, in the absence of the external driving protocol, a well defined partition function for the reduced system, from which the free energy, internal energy and the entropy can be introduced by taking the derivatives of the partition function. Alternatively, one may introduce the thermodynamic functions from the expectation values of suitable, physically sensible system operators. It has been argued that these two common practices, though totally compatible in the conventional weak-coupling thermodynamics, become inequivalent when the system-bath coupling is not vanishingly weak. The only temperature scale in that context is the initial temperature of the bath, so the free energy, the entropy and the Hamiltonian of mean force are introduced with respect to the initial bath temperature. In these earlier works, there is no natural way to bring in the effective or nonequilibrium temperature. Here, at least for a general Gaussian system, we can introduce the effective temperature via the nonequilibrium partition function, and from there define the nonequilibrium free energy and entropy. In particular, the entropy so defined gives the same results taken by both approaches~\cite{GT09,QTDpot}. When it comes to the internal energy, although it is still different from the expectation of the system's Hamiltonian through the whole evolution history, the internal energy we have defined here has an advantage over the internal energy defined with respect to the initial bath temperature. The latter, even for a simple case like the coupled Brownian oscillators, has been shown~\cite{{HA11}} to have negative heat capacity at the low temperature regime. The internal energy introduced in \eqref{E:gndbsds} is free from such anomalies because the hyper cotangent function is a monotonic function of positive effective temperature. In addition, it makes a consistent connection when quantum entanglement enters into the considerations, since the effective temperature may contain the same amount of information about the system-bath entanglement. All this will be discussed in the subsequent papers~\cite{HMF,QTD2}.

\section{Discussions}\label{S:rbtkfhskrt}

The four major findings of this paper as enumerated in the Introduction reflect also the main features of  nonequilibrium quantum  thermodynamics. For a dynamical Gaussian system strongly  coupled to a thermal bath studied here, we have shown from 1) a nonequilibrium partition function how 2) a nonequilibrium effective temperature and 3) a nonequilibrium free energy can be defined at all times in the system's history. These quantities enter in 4) a nonequilibrium thermodynamic relation $\mathcal{F}_{\textsc{s}}(t)=\mathcal{U}_{\textsc{s}}(t)-T_{\textsc{eff}} (t)\,\mathcal{S}_{vN}(t)$ where $\mathcal{U}_{\textsc{s}}$ is the internal energy and $\mathcal{S}_{vN}(t)$ is the von Neumann entropy of the reduced system. It has the same form as in the conventional (weak coupling) equilibrium thermodynamics, but each entry in this relation is now dynamical, valid at any moment of time in the system's evolution under fully nonequilibrium conditions. In addition, the thermodynamic functions in this new framework have the same functional dependence on the \textit{effective} temperature as their counterparts in conventional weak-coupling thermodynamics do on the \textit{system} temperature. It provides a systematic and complete formulation of nonequilibrium quantum thermodynamics on a par with the familiar weak-coupling equilibrium thermodynamics. In this paper, we have focused more on the nonequilibrium partition function, the nonequilibrium effective temperature and the nonequilibrium free energy. The other two equally important quantities, the Hamiltonian of mean force entering in the system's internal energy and the von Neumann entropy which recognizes the presence of quantum entanglement, will be treated with greater details in two subsequent papers~\cite{HMF,QTD2}. We wish to end this paper with some added information about the concept and context of effective temperature as have been used to describe glassy, aging or complex systems and to address some fundamental nonequilibrium issues.   

It might be useful to begin with a quick refresh of what temperature means. In microcanonical ensembles for an isolated system with internal energy $\mathcal{U}$, the inverse temperature $\beta\equiv T^{-1} $ can be defined by $\beta = \partial \ln \Omega/ \partial\mathcal{U}$ where $\Omega$ is the number of accessible states (e.g.,~\cite{FR65}). Temperature becomes a more directly relatable notion in canonical ensembles describing systems in weak contact with a thermal bath. Many systems placed in contact with a heat bath, given long enough time, will relax and equilibrate. If the coupling is \textit{weak} enough, the system is small enough compared to the bath, and the temperature is not at absolute zero, then the system will settle into a thermal state at a temperature very close to that of the bath. In equilibrium when a fluctuation-dissipation relation exists, we can with confidence refer to the system as at a certain temperature, namely, at the initial bath temperature. This is our conceptual starting point in trying to make sense of what temperature means for a system in nonequilibrium evolution prior to equilibration. First let us mention situations which deviate not much from thermal equilibrium. 
The concept of effective temperature is used in a loose sense for systems which in the mean can be described by a temperature. A somewhat exotic example is assigning a time-dependent Unruh temperature for circular acceleration~\cite{UnruhCern} (at constant speed but changing direction) while strictly speaking the Unruh effect refers to constant linear acceleration~\cite{HJCapri}. It is also used for systems close to equilibrium such as in the linear response regime, even for non-equilibrium linear response theory~\cite{NELR}. 

Effective temperature is also invoked for nonequilibrium systems  but only in phases which evolve sufficiently slowly, with low entropy production. As suggested in Cugliandolo's review~\cite{EffTemp} the energy density, and more generally, the averages of one-time-dependent observables, converge to finite values. A common example is glassy systems with short-range interactions where all these features are satisfied in some solvable mean-field-like description. Central to the correct identification of an effective temperature for these systems is to correctly identify the relaxation time scales and to deal with each of them separately. With well separated timescales, as in aging glassy systems, equilibrium fluctuation-dissipation theorem applies at each scale with its own effective temperature. With mixed timescales, as for example in active or granular fluids or in turbulence, temperature is no more well-defined, the dynamical nature of fluctuations fully emerges~\cite{Gnoli}.  

Finally, whether fluctuation-dissipation relations can exist in far from equilibrium situations \cite{Corberi} is also   actively debated. In a suitably slow regime the existence of a fluctuation-dissipation relation is conceivable and would provide some justification for introducing an effective temperature for that regime. The question is, under what conditions can the response function $R$ be related to the correlation functions $C$ of the unperturbed systems? It has been shown that for Markovian dynamics this can be done, although the correlation functions contain more than what is in the equilibrium case \cite{Parisi,Zannetti}. The derivative of $\partial R/\partial C$ has been used as an effective temperature~\cite{CKP}. Generalized fluctuation-dissipation theorems have also been suggested for active matter~\cite{GFDT}.

In the backdrop of the broader scope described above, the appearance in our system (which ostensibly is not `complex' -- one way to measure complexity is the existence of multiple time scales) of a function which changes in time and approaches the well-defined temperature upon equilibration -- the effective temperature -- should not be too concerning. It just takes some time to get use to in dealing with nonequilibrium systems.  In fact we hope the attractive features  of this nonequilibrium effective temperature can outweigh its temporary lack of familiarity: We mention three here:  One is obvious, that it changes with time, and since it carries information of both the system and the interaction with its environment, we can see how the energy to entropy ratio changes as the system encounters the environment, negotiates its way, ending in equilibration with its environment. Second feature is  strong coupling.  Usually temperature is a valid concept only for a small system very weakly coupled to a large environment. Here with strong coupling one can still use the notion of temperature all the way toward equilibration. Finally, thanks to the Gaussian nature of our system and environment and their coupling, the findings here, albeit not exhaustive, are based on  exact solutions.  This helps to reveal important physical features which cannot be accessed by relying on commonplace perturbative or approximate methods such as weak coupling, or Markovian assumptions. It also  enables one to rule out erroneous  predictions drawn from applying wrong approximations for regimes beyond their capabilities. In future work we will use the theoretical framework and technical platform constructed here to explore the many basic issues of nonequilibrium quantum thermodynamics.  
{}\\

\noindent {\bf Acknowledgments} This work was developed when J.-T. H. visited the Maryland Center for Fundamental Physics at the University of Maryland and when B. L. H. visited the Institute of Physics, Academia Sinica and the National Center for Theoretical Sciences in Hsinchu, Taiwan.


\end{document}